\documentclass[%
 reprint,
 amsmath,amssymb,
 aps,
]{revtex4-1}

\usepackage{graphicx}
\usepackage{dcolumn}
\usepackage{bm}
\usepackage{graphicx}
\usepackage{etoolbox}
\usepackage{natbib}
\usepackage{caption}
\usepackage{subcaption}
\usepackage{epstopdf}
\usepackage{color,soul}
\DeclareGraphicsExtensions{.eps}
	
\begin{document}

\preprint{APS/123-QED}

\title{Amplitude- and Frequency-based Dispersion Patterns and Entropy}

\author{Hamed Azami and Javier Escudero}

\affiliation{%
 \ Institute for Digital Communications, School of Engineering, University of Edinburgh, Edinburgh, EH9 3FB, UK
}%


\begin{abstract}
Permutation patterns-based approaches, such as permutation entropy (PerEn), have been widely and successfully used to analyze data. However, these methods have two main shortcomings. First, when a series is symbolized based on permutation patterns, repetition as an unavoidable phenomenon in data is not took in to account. Second, they consider only the order of amplitude values and so, some information regarding the amplitude values themselves may be ignored. To address these deficiencies, we have very recently introduced dispersion patterns and subsequently, dispersion entropy (DispEn). In this paper, we investigate the effect of different linear and non-linear mapping approaches, used in the algorithm of DispEn, on the characterization of signals. We also inspect the sensitivity of different parameters of DispEn to noise. Moreover, we introduce frequency-based DispEn (FDispEn) as a measure to deal with only the frequency of time series. The results suggest that DispEn and FDispEn with the log-sigmoid mapping approach, unlike PerEn, can detect outliers. Furthermore, the original and frequency-based forbidden dispersion patterns are introduced to discriminate deterministic from stochastic time series. The computation times show that DispEn and FDispEn are considerably faster than PerEn. Finally, we find that DispEn and FDispEn outperform PerEn to distinguish various dynamics of biomedical signals. Due to their advantages over existing entropy methods, DispEn and FDispEn are expected to be widely used for the characterization of a wide variety of real-world time series.

\end{abstract}

\pacs{Valid PACS appear here}
\keywords{Non-linear Analysis}
\maketitle



\section{Introduction}
Searching for patterns in signals and images is a fundamental problem and has a long history \cite{christopher2006pattern}. A pattern denotes an ordered set of numbers, shapes, or other mathematical objects, arranged based on a rule. Elements of a given set are usually arranged by the concepts of permutation and combination \cite{biggs1979roots}. Combination means a way of selecting elements or objects of a given set in which the order of selection does not matter. However, the order of objects is usually a crucial characteristic of a pattern \cite{christopher2006pattern,biggs1979roots}. 

In contrast, the concept of permutation indicates an arrangement of the distinct elements or objects of a given set into some sequences or orders \cite{donald1999art,biggs1979roots}. Permutation patterns have been studied occasionally, often implicitly, for over a century, although this area has grown significantly in the last three decades \cite{steingrimsson2010generalized}.

However, the concept of permutation does not allow repetition. There is no doubt that repetition is an unavoidable phenomenon in data. Furthermore, permutation considers only the order of amplitude values and so, some information regarding the amplitudes may be ignored \cite{azami2016amplitude,fadlallah2013weighted}. To overcome these shortcomings, we have recently introduced dispersion patterns, taking into account repetitions \cite{rostaghi2016dispersion}. 

The probability of occurrence of each potential dispersion or permutation pattern makes a key role to define the entropy of signals \cite{bandt2002permutation,rostaghi2016dispersion,shannon2001mathematical}. Entropy is a powerful measure to quantify the irregularity or uncertainty of time series \cite{rostaghi2016dispersion,shannon2001mathematical}. Assume we have a probability distribution $\textbf{s}$ with $\textit{N}$ potential patterns ${s_{1},s_{2},\dots,s_{N} }$. Based on the Shannon's definition, the entropy of the distribution \textbf{s} is $-\sum_{k=1}^{N} Pr\{s_{k}\}\log(Pr\{s_{k}\})$, where $Pr\{s_{k}\}$ is the probability of occurrence of pattern $s_{k}$ \cite{shannon2001mathematical}. When all the probability values are equal, the maximum entropy occurs, while if one probability is certain and the others are impossible, the minimum entropy is achieved \cite{shannon2001mathematical,rostaghi2016dispersion}.

Over the past three decades, a number of entropy methods have been introduced, such as sample entropy (SampEn) \cite{richman2000physiological}, permutation entropy (PerEn) \cite{bandt2002permutation}, and dispersion entropy \cite{rostaghi2016dispersion}. DispEn and PerEn are based on the Shannon's entropy definition applied to dispersion or permutation patterns. DispEn has been shown to have similar behaviour to SampEn when dealing with real and synthetic signals \cite{rostaghi2016dispersion}. However, as SampEn is based on conditional entropy \cite{richman2000physiological}, we do not compare dispersion and permutation entropies with SampEn. Nevertheless, we evaluate the relationship between the parameters of DispEn and SampEn in Appendix.

PerEn which is based on the permutation patterns or order relations among amplitudes of a time series is another widely-used entropy method \cite{bandt2002permutation}. For detailed information about the algorithm of PerEn please see \cite{bandt2002permutation}. PerEn is conceptually simple, computationally quick, and its computation cost is O(\textit{N})

Nevertheless, it has two main deficiencies directly derived from the fact that it considers permutation patterns. First, the original PerEn assumes a signal has a continuous distribution, therefore equal values are rare and can be ignored by ranking them based on the order of their emergence. However, while dealing with digitized signals with lower resolution, it may not be rational to simply ignore them \cite{bian2012modified,zanin2012permutation}. Second, when a time series is symbolized based on the permutation patterns (Bandt-Pompe procedure), only the order of amplitude values is taken into account and some information with regard to the amplitudes may be ignored \cite{fadlallah2013weighted}. To alleviate the first deficiency, modified PerEn (MPerEn) based on mapping equal values into the same symbol was developed \cite{bian2012modified}. To address the second shortcoming, Fadlallah \textit{et al}, have recently proposed weighted PerEn (WPerEn) to weight the motif counts by statistics derived from the time series patterns \cite{fadlallah2013weighted}. However, none of MPerEn and WPerEn addresses both the shortcomings of PerEn at the same time. Although our recently introduced amplitude-aware PerEn (AAPerEn) takes into account both the shortcomings \cite{azami2016amplitude}, its further parameters should be tuned.

To deal with the aforementioned shortcomings of PerEn, MPerEn, WPerEn, and AAPerEn, we have very recently introduced DispEn based on dispersion patterns to quantify the irregularity of time series \cite{rostaghi2016dispersion}. Note that, as the original DispEn is based on the amplitude values of signals \cite{rostaghi2016dispersion}, it might also be referred to as amplitude-based DispEn, although we will only use the term DispEn for conciseness. The results showed that DispEn, unlike PerEn, MPerEn, and WPerEn, is sensitive to changes in simultaneous frequency and amplitude values and bandwidth of time series and that DispEn outperformed PerEn in terms of discrimination of diverse biomedical states \cite{rostaghi2016dispersion}. Furthermore, as DispEn needs to neither sort the amplitude values of each embedding vector nor calculate every distance between any two composite delay vectors with embedding dimensions \textit{m} and $m+1$, it is noticeably faster than PerEn, WPerEn, and AAPerEn \cite{rostaghi2016dispersion}.

In this article, we investigate the effect of different parameters and mapping algorithms on the ability of DispEn to quantify the irregularity or uncertainty of signals for the first time. Note that these issues were not the scope of our last paper, which developed DispEn \cite{rostaghi2016dispersion}. In this paper, we introduce frequency-based DispEn (FDispEn) to deal with only the frequency-based entropy patterns (features). We also introduce the concepts of forbidden amplitude- and frequency-based dispersion patterns and show that they can be used to distinguish deterministic from stochastic time series. To assess the DispEn and FDispEn, we use both synthetic and real datasets. 

\section{Methods}
In this section, we describe DispEn and FDispEn in detail.
\subsection{Dispersion Entropy (DispEn) with Different Mapping Techniques}
Given a univariate signal $\textbf{x}=\{\textit{x}_{1}, \textit{x}_{2},\dots, \textit{x}_{N}\}$ with length \textit{N}, the DispEn algorithm is as follows:

1) First, $x_{j}(j=1,2,\dots,N)$ are mapped to $\textit{c}$ classes with integer indices from 1 to $\textit{c}$. The classified  signal is $u_{j}(j=1,2,\dots,N)$. A number of linear and non-linear mapping techniques, introduced later, can be used in this step.

2) Time series $\textbf{u}^{m,c}_i$ are made with embedding dimension $m$ and time delay $d$ according to $\textbf{u}^{m,c}_i=\{u^{c}_i,{u}^{c}_{i+d},\\\dots,{u}^{c}_{i+(m-1)d}\} $, $i=1,2,\dots,N-(m-1)d$ \cite{rostaghi2016dispersion,bandt2002permutation}. Each time series $\textbf{u}^{m,c}_i$ is mapped to a dispersion pattern ${{\pi }_{{{v}_{0}}{{v}_{1}}\dots{{v}_{m-1}}}}$, where $u_{i}^{c}=v_0$, $u_{i+d}^{c}=v_1$,\dots, $u_{i+(m-1)d}^{c}=v_{m-1}$. The number of possible dispersion patterns assigned to each vector $\textbf{u}^{m,c}_i$ is equal to $c^{m}$, since the signal has $m$ members and each member can be one of the integers from 1 to $c$ \cite{rostaghi2016dispersion}. 

3) For each of $c^{m}$ potential dispersion patterns ${{\pi }_{{{v}_{0}}\dots{{v}_{m-1}}}}$, relative frequency is obtained as follows:
\begin{equation}
	\begin{split}
		p({{\pi }_{{{v}_{0}}\dots{{v}_{m-1}}}})=\\
		\frac{\#\{i\left| i\le N-(m-1)d,\mathbf{u}_{i}^{m,c}\text{ has type }{{\pi }_{{{v}_{0}}\dots{{v}_{m-1}}}} \right.\}}{N-(m-1)d}
	\end{split}
\end{equation}

\setlength{\parindent}{0pt}%
where $\#$ means cardinality. In fact, $p({{\pi }_{{{v}_{0}}\dots{{v}_{m-1}}}})$ shows the number of dispersion patterns of ${{\pi }_{{{v}_{0}}\dots{{v}_{m-1}}}}$ that is assigned to $\textbf{u}^{m,c}_i$, divided by the total number of embedded signals with embedding dimension $m$.
\setlength{\parindent}{9pt}

4) Finally, based on the Shannon's definition of entropy, the DispEn value is calculated as follows:
\begin{equation}
	\begin{split}
		DispEn(\mathbf{x},m,c,d)=\\
		-\sum_{\pi=1}^{c^m} {p({{\pi}_{{{v}_{0}}\dots{{v}_{m-1}}}})\cdot\ln }\left(p({{\pi}_{{{v}_{0}}\dots{{v}_{m-1}}}}) \right)      
	\end{split}
\end{equation}

As an example, let's have a series $x=\{3.6,4.2,1.2,3.1,4.2,2.1,3.3,4.6,6.8,8.4\}$, shown on the top left of Figure 1. We want to calculate the DispEn value of \textbf{x}. For simplicity, we set $d=1$, $m=2$, and $c=3$. The $3^2=9$ potential dispersion patterns are depicted on the right of Figure 1. $x_j$ ($j=1,2,\dots,10$) are linearly mapped into 3 classes with integer indices from 1 to 3, as can be seen in Figure 1. Next, a window with length 2 (embedding dimension) moves along the signal and the number of each of dispersion patterns is counted. The relative frequency is shown on the bottom left of Figure 1. Finally, using Eq. 2, the DispEn value of \textbf{x} is equal to $-(\frac{2}{9}\ln(\frac{2}{9})+\frac{2}{9}\ln(\frac{2}{9})+\frac{2}{9}\ln(\frac{2}{9})+\frac{1}{9}\ln(\frac{1}{9})+\frac{1}{9}\ln(\frac{1}{9})+\frac{1}{9}\ln(\frac{1}{9}))=1.7351$.

\begin{figure*}
	\centering
	\includegraphics[width=19cm,height=8cm]{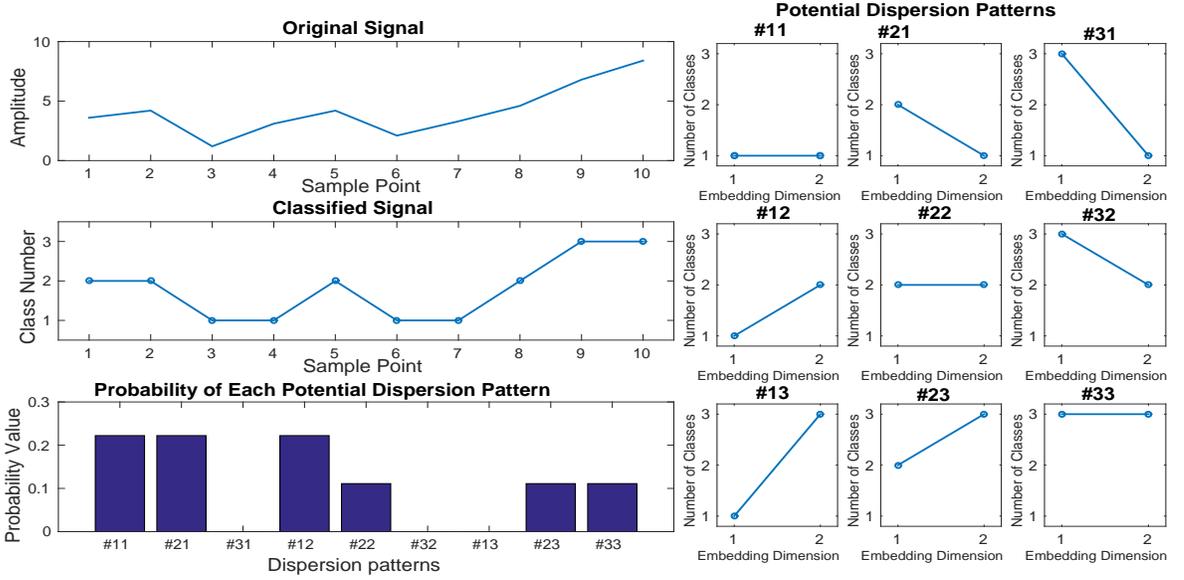}
	\caption{Illustration of the DispEn algorithm using linear mapping of $x=\{3.6,4.2,1.2,3.1,4.2,2.1,3.3,4.6,6.8,8.4\}$ with the number of classes 3 and embedding dimension 2. }
	\label{figurelabel}
\end{figure*}

If all possible dispersion patterns have equal probability value, the DispEn reaches to its highest value, which has a value of $ln ({{c}^{m}})$. In contrast, when there is only one $ p({{\pi }_{{{v}_{0}}\dots{{v}_{m-1}}}}) $ different from zero, which demonstrates a completely regular/predictable time series, the smallest value of DispEn is obtained \cite{rostaghi2016dispersion}. Note that we use the normalized DispEn as $\frac{DispEn}{\ln ({{c}^{m}}) }$ in this study \cite{rostaghi2016dispersion}.

\subsection{Frequency-based Dispersion Entropy (FDispEn)}
When only the frequency of a signal is relevant, there is no difference between dispersion patterns $\{1,3,4\}$ and $\{2,4,5\}$ or $\{1,1,1\}$ and $\{3,3,3\}$. To take into account only the frequency of signals, we introduce FDispEn in this article. In fact, FDispEn considers the differences between adjacent elements of dispersion patterns, termed frequency-based dispersion patterns. In this way, we have vectors with length $m-1$ which each of their elements changes from $-c+1$ to $c-1$. Thus, there are $(2c-1)^{m-1}$ potential frequency-based dispersion patterns. The only difference between DispEn and FDispEn algorithms is the potential patterns used in these two approaches.

As an example, let's have a signal $x=\{3,4.5,6.2,5.1,3.2,1.2,3.5,5.6,4.9,8.4\}$. We set $d=1$, $m=3$, and $c=2$, leading to have $3^2=9$ potential frequency-based dispersion patterns ($\{(-1,-1),(-1,0),(-1,1),(0,-1),(0,0),\\(0,1),(1,-1),(1,0),(1,1)\}$). Then, $x_j$ ($j=1,2,\dots,10$) are linearly mapped into 2 classes with integer indices from 1 to 2 ($\{1,1,2,2,1,1,1,2,2,2\}$). Afterwards, a window with length 3 moves along the time series and the differences between adjacent elements are calculated ($\{(0,1),(1,0),(0,-1),(-1,0),(0,0),(0,1),(1,0),(0,0)\}$). Afterwards, the number of each frequency-based dispersion pattern is counted. Finally, using Eq. 2, the DispEn value of \textbf{x} is equal to $-(\frac{1}{8}\ln(\frac{1}{8})+\frac{1}{8}\ln(\frac{1}{8})+\frac{2}{8}\ln(\frac{2}{8})+\frac{2}{8}\ln(\frac{2}{8})+\frac{2}{8}\ln(\frac{2}{8}))=1.5596$.  

\subsection{Mapping Approaches used in DispEn and FDispEn}

A number of linear and non-linear methods can be used to map the original signal $x_{j}(j=1,2,\dots,N)$ to the classified signal $u_{j}(j=1,2,\dots,N)$. The simplest and fastest algorithm is the linear mapping. However, when maximum or minimum values are noticeable larger or smaller than the mean/median value of the signal, the majority of $x_{j}$ are mapped to only few classes. To alleviate the problem, we can sort $x_{j}(j=1,2,\dots,N)$ and then divide them into \textit{c} classes in which each of them includes equal number of $x_j$.

We also use several non-linear mapping techniques. Many natural processes show a progression from small beginnings that accelerates and approaches a climax over time (e.g., a sigmoid function) \cite{tuffery2011data}. When there is not a detailed description, a sigmoid function is frequently used \cite{gibbs2000variational}. Well-known log-sigmoid (logsig) and tan-sigmoid (tansig) transfer functions are respectively defined as:

\begin{equation}
	y_j=\frac{1}{e^{-\frac{x_j-\mu}{\sigma}}}
\end{equation}

\begin{equation}
	y_j=\frac{2}{1+e^{-2\frac{x_j-\mu}{\sigma}}}-1
\end{equation}
where $\sigma$ and $\mu$ are the standard deviation (SD) and mean of time series \textbf{x}, respectively.

The cumulative distribution functions (CDFs) for many common probability distributions are sigmoidal. The most well-known such example is the error function, which is related to the CDF of a normal distribution, termed normal CDF (NCDF). NCDF of \textbf{x} is calculated as follows:

\begin{equation}
	y_{j}=\frac{1}{\sigma\sqrt{2\pi}}\int\limits_{-\infty}^{x_{j}}{e}^{\frac{-(t-\mu)^2}{2\sigma^2}}\,\mathrm{d}t\end{equation}

Each of the aforementioned techniques maps $\textbf{x}$ into $\textbf{y}=\{\textit{y}_{1}, \textit{y}_{2},\dots, \textit{y}_{N}\}$, ranged from $\alpha$ to $\beta$. Then, we use a linear algorithm to assign each $y_{j}$ to a real number $z_{j}$ from 0.5 to $c+0.5$. Then, for each member of the mapped signal, we use $u_{j}^{c}=\mbox{round}(z_{j})$, where  $u_{j}^{c}$ denotes the $j^{th}$ member of the classified signal and rounding involves either increasing or decreasing a number to the next digit \cite{rostaghi2016dispersion}.

\section{Parameters of DispEn, FDispEn, and PerEn}
\begin{figure*}
	\centering
	\includegraphics[width=19cm,height=10cm]{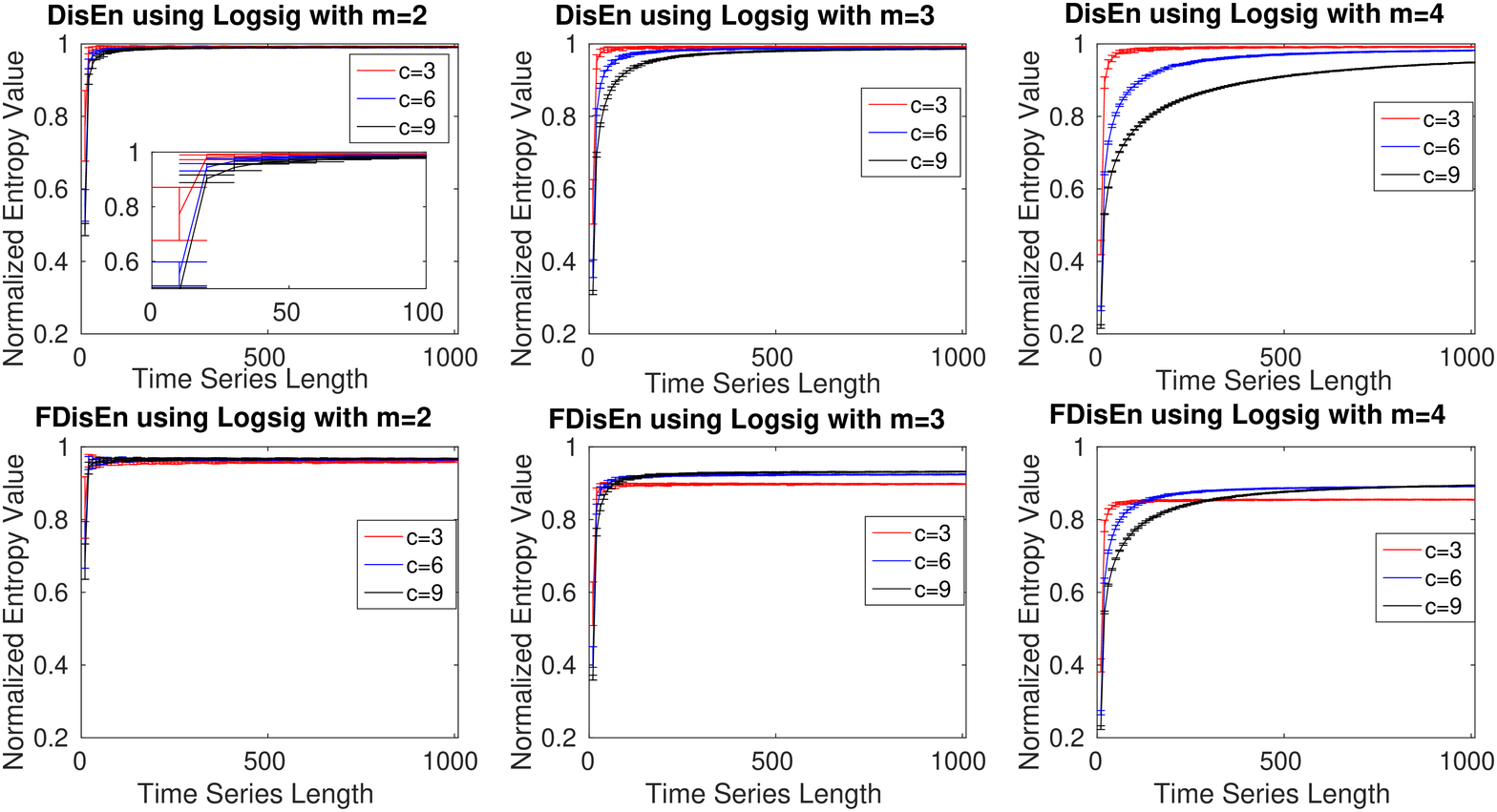}
	\caption{Mean and SD of results obtained by the DispEn and FDispEn with logsig and different values of embedding dimension and number of classes for 40 realizations of univariate white noise.}
	\label{Parameters_of_Dispersion_Entropy}
\end{figure*}

To assess the sensitivity of DispEn and FDispEn with logsig, and PerEn to the signal length, embedding dimension \textit{m}, and number of classes \textit{c}, we use 40 realizations of univariate white noise. Note that we will show why logsig is an appropriate mapping technique for DispEn and FDispEn to characterize signals. The mean and SD of results, depicted in Figure~\ref{Parameters_of_Dispersion_Entropy}, show that DispEn and FDispEn need a smaller number of sample points to reach to their maximum values for a smaller number of classes or smaller embedding dimension. This is in agreement with the fact that we need at least $\ln(c^m)$ \cite{rostaghi2016dispersion} and $\ln((2c-1)^{m-1})$ sample points to reach the maximum value of DispEn and FDispEn, respectively. The profiles also suggest that the greater the number of sample points, the more robust DispEn estimates, as seen from the errorbars.

We also inspect the relationship between noise power levels and DispEn with different number of classes. To this end, we use a logistic map added with different levels of noise power. This analysis is dependent on the model parameter $\alpha$ as: $x_{j}=\alpha x_{j-1}(1-x_{j-1})$, where the signal $x$ was generated with the parameter $\alpha$ varying from 3.5 to 3.99. In case $\alpha$ equals to 3.5, the time series oscillates among four values. When $3.5<\alpha <3.57$, the signal is periodic and the number of values doubles progressively. For $3.57\le\alpha\le3.99$, the series is chaotic, albeit it has segments with periodic behaviour (e.g., $\alpha\approx3.8$) \cite{baker1996chaotic,ferrario2006comparison,escudero2009interpretation}. The length and sampling frequency of the signal are respectively 100 s and 150 Hz. We added 40 independent realizations of white Gaussian noise (WGN) with different signal-to-noise ratios (SNRs) per sample, ranging from 0 to 50 dB, to the logistic map. We then employed a sliding window with length 1500 sample points and $50\%$ overlap moving along the signal to show the effect of noise power on each segment (window) of the signal. To compare the sensitivity of each method to WGN, we calculate \textit{NrmEntN} as the entropy value of each segment with noise over the entropy value of its corresponding segment without noise ($NrmEntN=\frac{\texttt{entropy of a series with noise}}{\texttt{entropy of a series without noise}}$).

\begin{figure*}
	\centering
	\includegraphics[width=19cm,height=7cm]{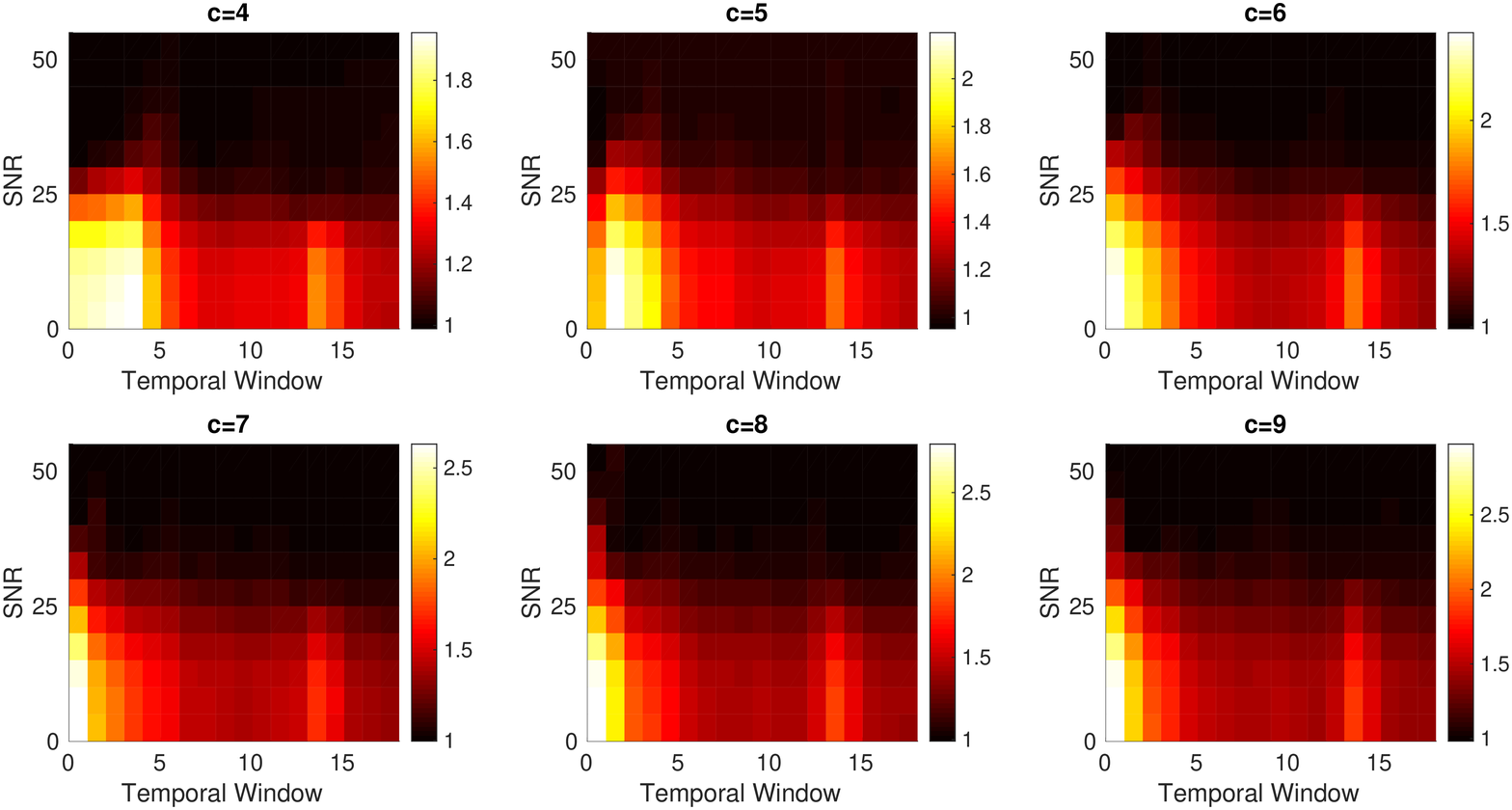}
	\small{(a) Average}
	\includegraphics[width=19cm,height=7cm]{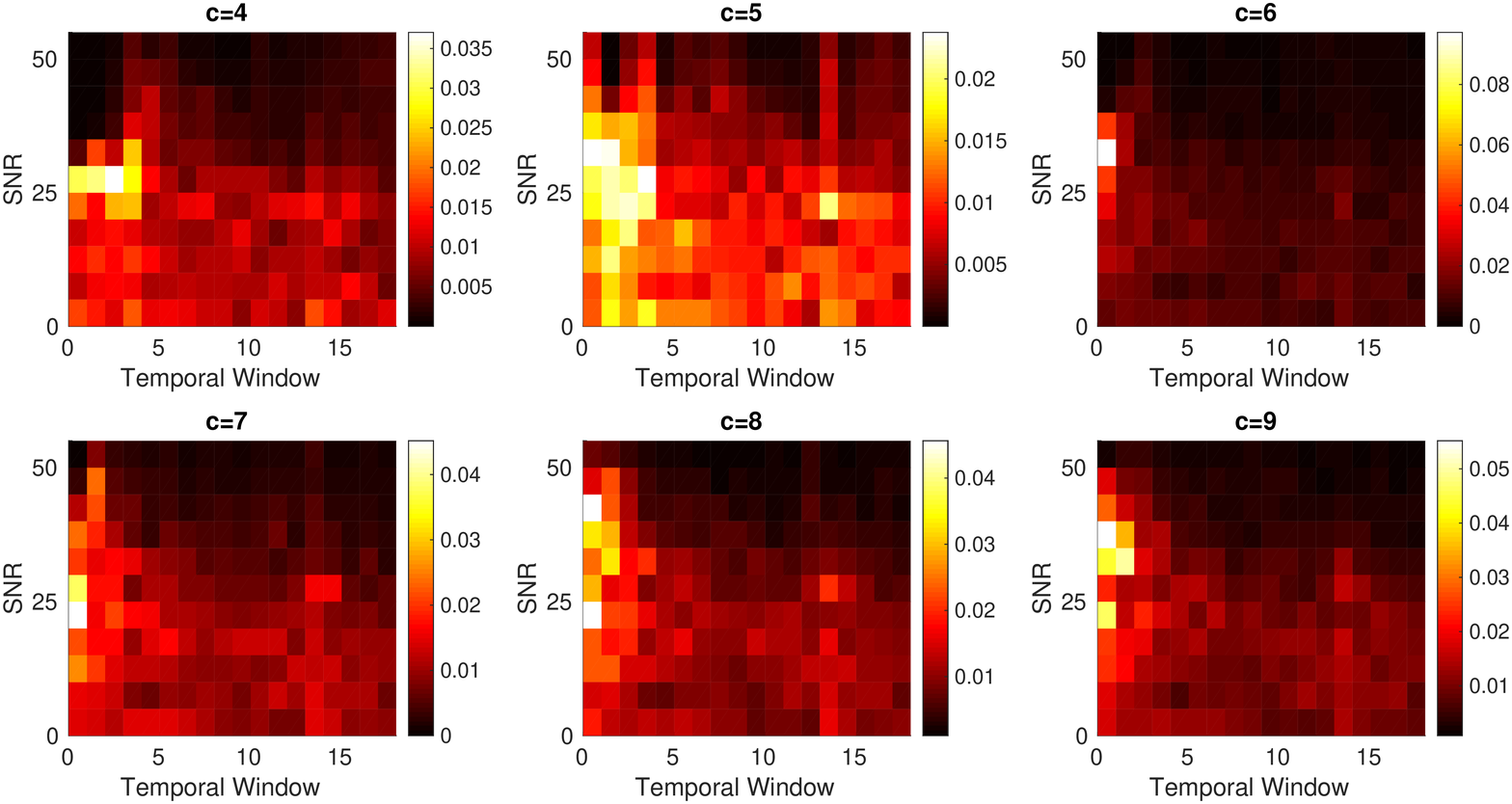}
	\small{(b) SD}
	\caption{(a) Average and (b) SD of $NrmEntN=\frac{\texttt{entropy of a series with noise}}{\texttt{entropy of a series without noise}}$ values obtained by the DispEn using logsig with different number of classes computed from the logistic map with additive 40 independent realizations of WGN with different noise power. \textit{NrmEntN} compares the sensitivity of DispEn to WGN with different SNRs. We used a window with length 1500 samples moving along the logistic map (temporal window) with varying parameter $\alpha$ from 3.5 to 3.99 showing an increase in entropy values along the signal, except for the downward spikes in the windows of periodic behavior (e.g., for $\alpha=3.8$). Darker means better results in this kind of figures.}
	\label{Average_Sensivity_SNR_noise_DisEn_Class}
\end{figure*}

The average and SD values of results obtained by the DispEn using logsig with different number of classes computed from the logistic map whose parameter ($\alpha$) varies from 3.5 to 3.99 with additive 40 independent realizations of WGN with different noise powers are shown in Figure~\ref{Average_Sensivity_SNR_noise_DisEn_Class}(a) and~\ref{Average_Sensivity_SNR_noise_DisEn_Class}(b), respectively. We set $m=2$ for DispEn \cite{rostaghi2016dispersion}. Figure~\ref{Average_Sensivity_SNR_noise_DisEn_Class} suggests that the larger the number of classes, the larger the value of \textit{NrmEntN}, as expected. Thus, when dealing with a low SNR, it is recommended to have a small \textit{c}. In fact, when \textit{c} is too large, small change may alter the class of a sample and therefore, the DispEn method might be sensitive to noise. On the other hand, if SNR is too large, we can choose a large \textit{c}. When \textit{c} is too small, two amplitude values that are far from each other may be assigned to a similar class, leading to unreliable entropy values. Thus, we need to have a trade-off between large and small number of classes c. As the SD and average of results are appropriate for $c=6$ (Figure~\ref{Average_Sensivity_SNR_noise_DisEn_Class}) and for simplicity, we set $c=6$ for all the simulations below.

Compared with DispEn, in the FDispEn algorithm, we have vectors with length $m-1$ which each of their elements changes from $-c+1$ to $c-1$. Thus, we set $m=3$ here. Similar what we did for DispEn, we changed \textit{c} from 4 to 9 for FDispEn. We found that $c=5$ leads to stable results when dealing with noise (results are not shown herein). Thus, we set $c=5$ for all simulations using FDispEn, although the range $2<c<9$ results in similar profiles. 

Overall, the parameter \textit{c} is chosen to balance the quality of entropy estimates with the loss of signal information. To avoid the impact of noise on signals, a small \textit{c} is recommended. In contrast, for a small \textit{c}, too much detailed data information is lost, leading to poor probability estimates. Thus, a trade-off between large and small \textit{c} values is needed.

\section{Evaluation of Mapping Approaches for DispEn and FDispEn}

\begin{figure*}
	\centering
	\includegraphics[width=19cm,height=10cm]{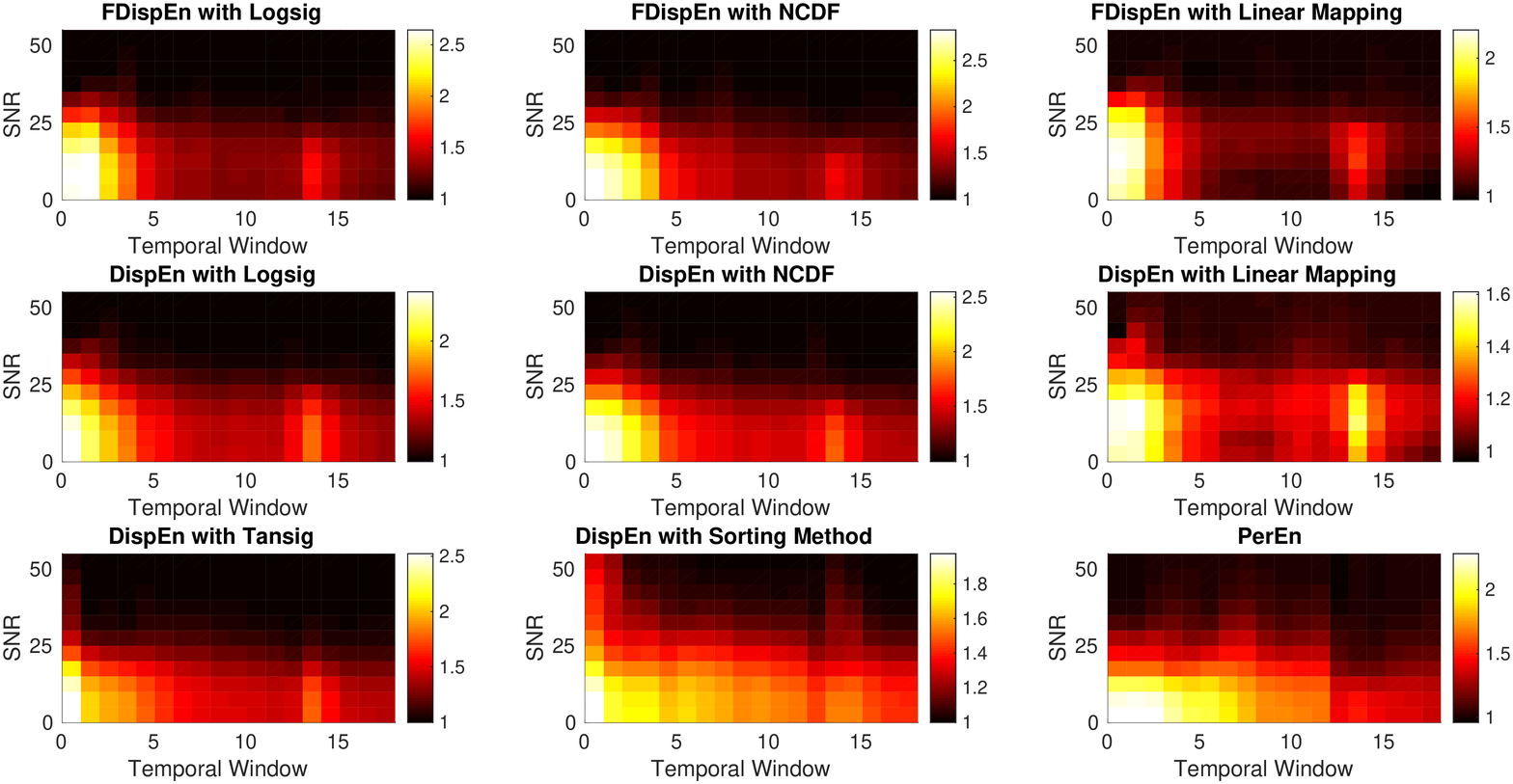}
	\small{(a) Average}
	\includegraphics[width=19cm,height=10cm]{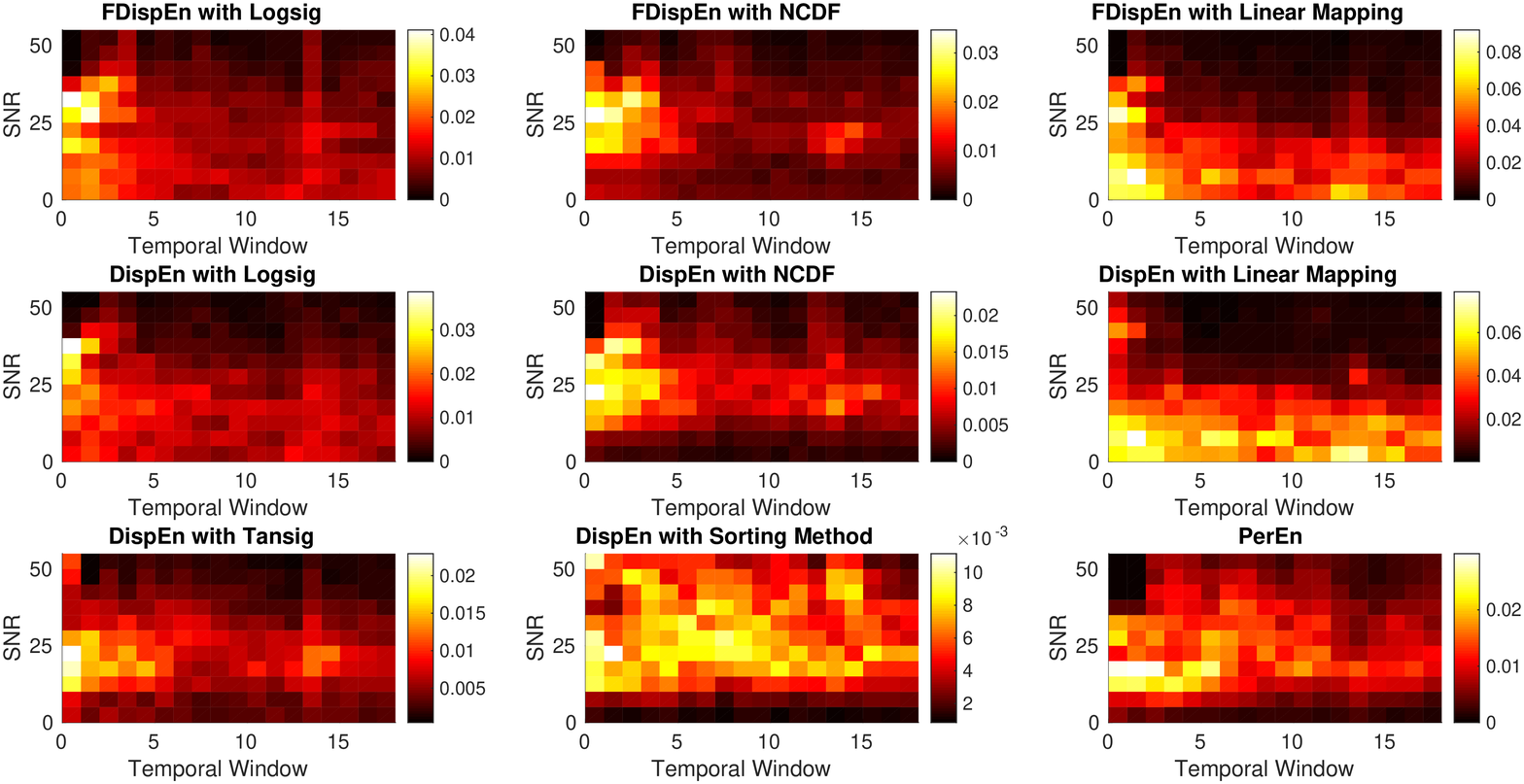}
	\small{(b) SD}
	\caption{(a) Average and (b) SD of $NrmEntN=\frac{\texttt{entropy of a series with noise}}{\texttt{entropy of a series without noise}}$ values obtained by the DispEn and FDispEn with different mapping techniques computed from the logistic map with additive 40 independent realizations of WGN with different noise power. \textit{NrmEntN} compares the sensitivity of each method to WGN with different SNRs. We used a window with length 1500 samples moving along the logistic map (temporal window) with varying parameter $\alpha$ from 3.5 to 3.99 showing an increase in entropy values along the signal, except for the downward spikes in the windows of periodic behavior (e.g., for $alpha=3.8$). Darker means better results in this kind of figures.}
	\label{Sensivity_SNR_Entropy_diff_mapping}
\end{figure*}

To evaluate the ability of DispEn and FDispEn with different mapping techniques to distinguish changes from periodicity to non-periodic non-linearity with different levels of noise, the described logistic map with additive noise is used. The average and SD of results obtained by the DispEn and FDispEn with different mapping techniques, and
PerEn are depicted in Figure~\ref{Sensivity_SNR_Entropy_diff_mapping}(a) and ~\ref{Sensivity_SNR_Entropy_diff_mapping}(b), respectively. The entropy values of the logistic map generally increase along the signal, except for the downward spikes in the windows of periodic behavior (e.g., for $\alpha=3.8$), in agreement with Figure 4.10 (page 87 in \cite{baker1996chaotic}) and previous studies \cite{azami2017refineddd,escudero2009interpretation}. As noise affects more on periodic oscillations, \textit{NrmEntN} is larger at lower temporal scales. The range of mean values show that DispEn and FDispEn with different mapping algorithms, and PerEn are similar, while dealing with the different levels of noise power. Figure~\ref{Sensivity_SNR_Entropy_diff_mapping}(b) suggests that when all signals have equal SNR values, the entropy values are stable for all the methods.

The ranges of mean values show that DispEn with sorting method and linear mapping lead to the most stable results. Although DispEn with sorting method, unlike PerEn, takes into account repetitions, it considers only the order of amplitude values and thus, some information regarding the amplitudes may be discarded. For instance, as evidenced later, DispEn with sorting method cannot detect the outliers or spikes which is noticeably larger or smaller than their adjacent values. For DispEn with linear mapping, when maximum or minimum values are noticeable larger or smaller than the mean/median value of the signal, the majority of $x_{j}$ are mapped to only few classes \cite{rostaghi2016dispersion}. Thus, for simplicity, we use DispEn and FDispEn with logsig for all the simulations below.

Noise is frequently considered as an unwanted component or disturbance to a system or data, whereas recent studies have shown that noise can play a beneficial role in systems \cite{cohen2005history,sejdic2013necessity}. In any case, it has been evidenced that noise is an essential ingredient in the systems and has a noticeable effect on many aspects of science and technology, such as engineering, medicine, and biology \cite{cohen2005history,sejdic2013necessity}. White, pink, and brown noise are three well-known unavoidable noise signals in the real world. White noise is a random signal having equal energy across all frequencies. The power spectral density of white noise is as $S(f)=C_w$, where $C_w$ is a constant \cite{sejdic2013necessity}. Pink and brown noise are random processes suitable for modelling evolutionary or developmental systems \cite{keshner19821}. The power spectral density $S(f)$ of pink and brown noise are as $\frac{C_p}{f}$ and $\frac{C_b}{f^2}$, respectively, where $C_p$ and $C_b$ are constants \cite{sejdic2013necessity,keshner19821}.

To evaluate the ability of DispEn and FDispEn methods with different mapping algorithms, and PerEn to distinguish the dynamics of different noise signals, we created 40 realizations of white, brown, and pink noise signals with different lengths changing from 10 to 1000 sample points. Note that, as the maximum value of PerEn is $\ln(m!)$ \cite{azami2015evaluation}, we use normalized PerEn as $\frac{PerEn}{\ln(m!)}$ in this study. We set $m=4$ for PerEn \cite{kowalski2007bandt}, $m=2$ and $c=6$ for DispEn \cite{rostaghi2016dispersion}, and $m=3$ and $c=5$ for FDispEn as recommended before.

\begin{figure*}
	\centering
	\includegraphics[width=19cm,height=12cm]{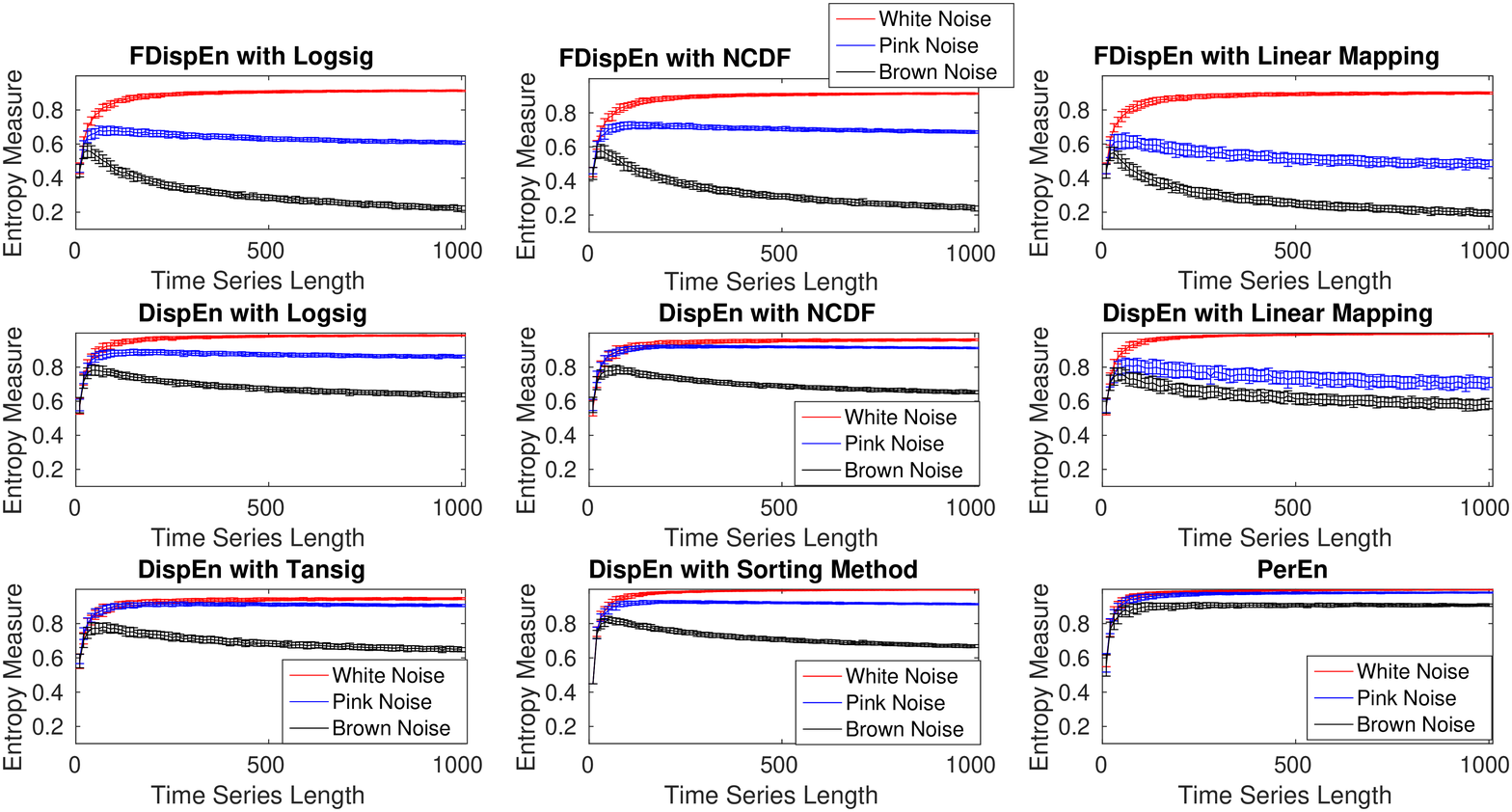}
	\caption{Mean and SD of entropy values obtained by DispEn and FDispEn with different mapping techniques and PerEn, computed from 40 different white, brown, and pink noise signals.}
	\label{Different_Noise_length}
\end{figure*}

Figure~\ref{Different_Noise_length} shows that DispEn and FDispEn with different mapping approaches distinguish brown, pink, and white noise series with different lengths. Their results are in agreement with the fact that white noise is the most irregular signal, followed by pink and brown noise, in that order, based on the power spectral density of white, pink, and brown noise \cite{cohen2005history,sejdic2013necessity}. However, there are some overlaps between the DispEn with tansig, and PerEn values for short pink and white noise time series, suggesting a superiority of DispEn and FDispEn with different mapping approaches, except tansig, over PerEn.

\section{Computation Cost of DispEn and FDispEn}
In order to assess the computational time of DispEn and FDispEn with logsig, compared with PerEn, we use random time series with different lengths, changing from 300 to 100,000 sample points. The results are depicted in Table~\ref{Results_Computation_Cost}. The simulations have been carried out using a PC with Intel (R) Xeon (R) CPU, E5420, 2.5 GHz and 8-GB RAM by MATLAB R2015a. 


For DispEn, FDispEn, and PerEn, the larger the \textit{m} value, the longer the computation time. For short signals, the differences between the computation time values for DispEn and FDispEn, and PerEn are not considerable. However, for long signals, DispEn is relatively faster than FDispEn and the latter is noticeably quicker than PerEn. This is in agreement with the fact that DispEn and FDispEn, unlike PerEn, do not need to sort each of the embedded series.

\begin{table*}
	\centering
	\setlength{\tabcolsep}{7pt}
	\label{tab:table4}\caption{Computational time of DispEn and FDispEn with logsig, and PerEn with different embedding dimension values.} 
	\begin{tabular}{c*{8}{c}}
		Number of samples $\rightarrow$ & 300   & 1,000 & 3,000 & 10,000& 30,000 &100,000 &300,000 \\
		\hline
		DispEn ($m=2$) &     0.0018 s& 0.0019 s&0.0026 s& 0.0041 s& 0.0078 s&0.0202 s&0.0606 s\\
		DispEn ($m=3$) &       0.0026 s&   0.0032 s&  0.0050 s&   0.0106 s &  0.0268 s&  0.0734 s&0.2296 s\\
		DispEn ($m=4$) &0.0071 s&  0.0091 s  & 0.0165 s  & 0.0411 s&   0.1199 s&   0.3644 s& 1.0850 s \\
		DispEn ($m=5$)  & 0.0307 s  & 0.0394 s&   0.0671 s &  0.1740 s  & 0.5650 s&   2.1301 s& 6.2143 s \\
		FDispEn ($m=2$) &    0.0026 s&   0.0026 s  & 0.0028 s&   0.0036 s &  0.0063 s & 0.0142 s&0.0481 s \\
		FDispEn ($m=3$) &      0.0033 s  & 0.0033 s&   0.0038 s&   0.0071 s &  0.0171 s &  0.0469 s& 0.1119 s \\
		FDispEn ($m=4$)   &      0.0071 s&   0.0088 s  & 0.0150 s &  0.0380 s &  0.1012 s &  0.3179 s& 0.8654 s \\
		FDispEn ($m=5$)   &       0.0562 s &  0.0710 s&   0.1195 s &  0.3027 s&   0.8896 s&   2.9412 s& 7.899 s\\
		PerEn ($m=2$) &  0.0042 s&   0.0109 s &  0.0266 s&   0.0884 s&   0.2396 s &  0.7962 s& 2.8734 s \\
		PerEn ($m=3$) &   0.0055 s &  0.0117 s&   0.0347 s &  0.1025 s&   0.3088 s &  1.0253 s& 4.3009 s \\
		PerEn ($m=4$) &    0.0068 s &  0.0227 s&   0.0667 s &  0.2235 s&   0.6657 s &  2.2280 s& 10.5902
		s \\
		PerEn ($m=5$)  &  0.0247 s&   0.0820 s&   0.2469 s&   0.8218 s&   2.4601 s  & 8.2029 s& 44.7658 s\\  
	\end{tabular}
	\label{Results_Computation_Cost}
\end{table*}

\section{Forbidden Amplitude- and Frequency-based Dispersion Patterns}
In this section, we introduce forbidden amplitude- and frequency-based dispersion patterns and explore the use of theses concepts to discriminate deterministic from stochastic time series. Forbidden patterns denote those patterns that do not appear at all in the analysed signal \cite{carpi2010missing,zanin2012permutation}. There are two reasons behind the existence of forbidden patterns. First, a signal with finite length do not have a number of potential patterns (false forbidden patterns). For example, the time series $\{1,2,3,2.1,1,4\}$ has only 4 permutations from 6 potential permutation patterns with $m=3$. Thus, the permutations $\{231\}$ and $\{312\}$ can be considered as false forbidden patterns. The second reason is based on the dynamical nature of the systems creating a signal. When signals made by an unconstrained stochastic process, all possible permutation patterns are appeared and there is no forbidden pattern. In contrast, it was evidenced that deterministic one-dimensional maps always have forbidden permutation or ordinal patterns \cite{carpi2010missing,amigo2007true}.

\begin{figure}
	\centering
	\includegraphics[width=9cm,height=6cm]{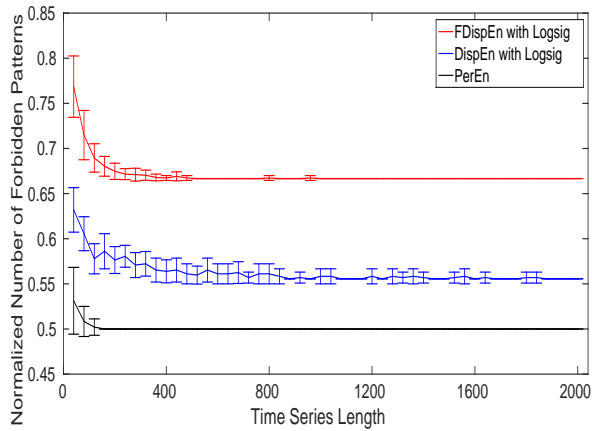}
	\caption{Mean and SD of the normalized number of forbidden amplitude- and frequency-based dispersion and permutation patterns ($\frac{\texttt{number of forbidden patterns}}{\texttt{potential number of patterns}}$) as functions of the signal length.}
	\label{Forbidden_Dispersion_Patterns}
\end{figure}

Theoretically, as permutation patterns are a subset of amplitude- and frequency-based dispersion patterns \cite{rostaghi2016dispersion}, the two latter patterns can discriminate deterministic from stochastic time series too. The normalized number of forbidden (missing) dispersion and permutation patterns as a function of the signal length using the logistic map $x_{t+1}=4x_t(1-x_t)$ \cite{amigo2007true} for DispEn and FDispEn with logsig, and PerEn are shown in Figure~\ref{Forbidden_Dispersion_Patterns}. Note that the normalized number of forbidden patterns is equal to the number of forbidden patterns over the potential number of patterns ($m!$, $c^m$, and $(2c-1)^{m-1}$ for respectively PerEn, DispEn, and FDispEn). As can be seen in Figure~\ref{Forbidden_Dispersion_Patterns}, for short signals we have a number of false forbidden patterns. The results evidence that more than half of the dispersion and permutation patterns are forbidden. On the whole, the results show that both the amplitude- and frequency-based dispersion patterns can be used to differentiate deterministic from stochastic time series.

\section{Application of DispEn and FDispEn to Outlier Detection}
\begin{figure}
	\centering
	\includegraphics[width=9cm,height=6cm]{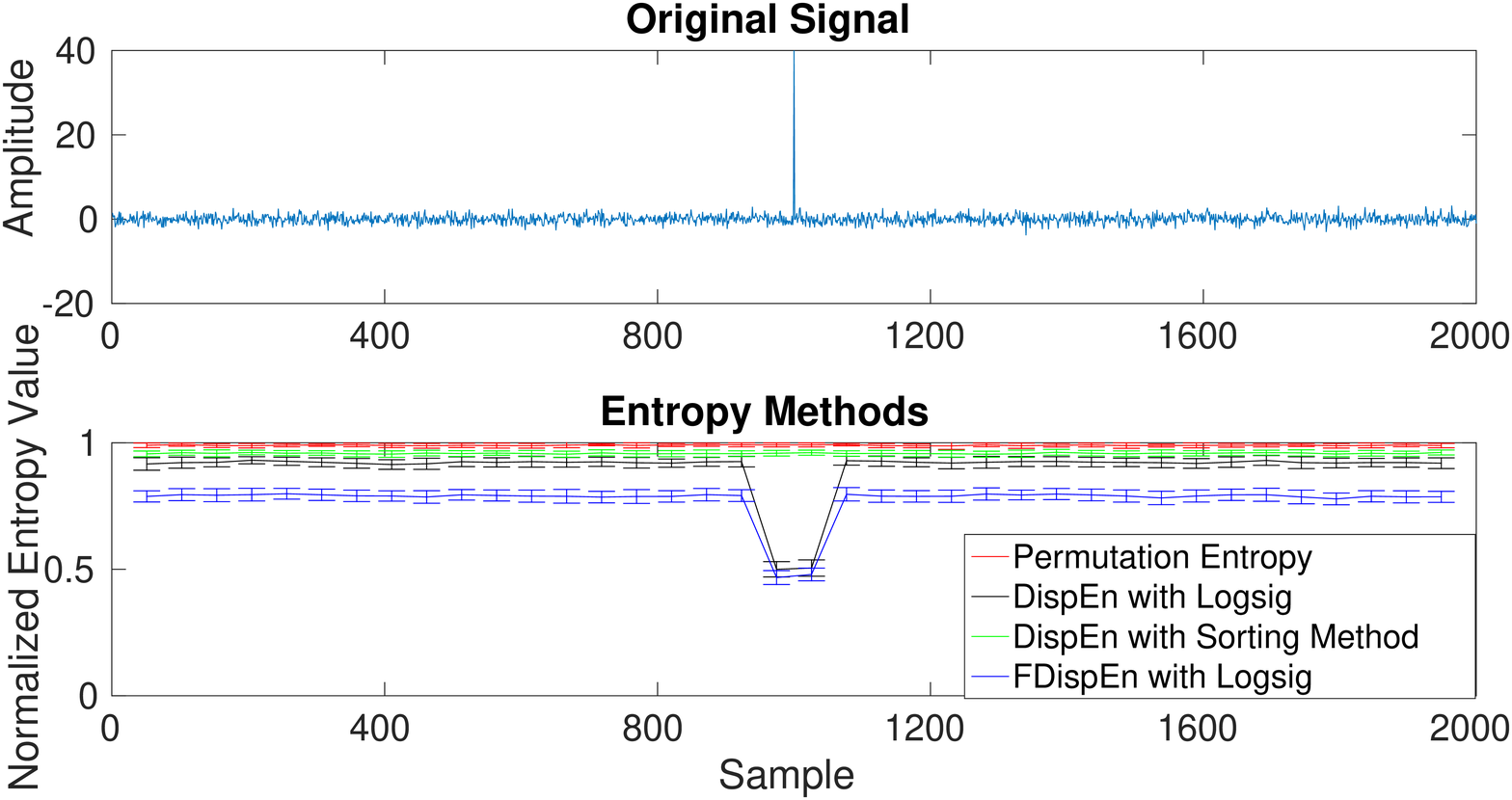}
	\caption{Mean and SD of results obtained by PerEn, and DispEn and FDispEn with logsig computed from 40 different synthetic test signals for spike detection \cite{fadlallah2013weighted,azami2016amplitude}.}
	\label{Results_Spike_Detection}
\end{figure}

A key shortcoming of PerEn is its inability to discriminate distinct patterns of a certain motif and the sensitivity of patterns close to the noise \cite{fadlallah2013weighted,azami2016amplitude}. To this end, we suggest analyzing the behavior of PerEn, DispEn with logsig and sorting method, and FDispEn with logsig in the existence of 40 different realizations of an impulsive and noise time series \cite{fadlallah2013weighted,azami2016amplitude}. Figure~\ref{Results_Spike_Detection} depicts 2000 sample points of a signal including an impulse and additive WGN.

A window with length 100 samples and overlap of $90\%$ moves along the signal and the entropy values each segment (window) are calculated. We set $m=4$ for PerEn \cite{kowalski2007bandt}, $m=2$ for DispEn, and $m=3$ for FDispEn. The SD and mean of results are shown in Figure~\ref{Results_Spike_Detection}. Considerable decreases in the values of DispEn and FDispEn with logsig, unlike PerEn and DispEn with sorting method, are found in the impulse region. Although DispEn with sorting method has a good performance when dealing with noise (Figures~\ref{Different_Noise_length} and~\ref{Sensivity_SNR_Entropy_diff_mapping}), this technique cannot detect the dynamics of spikes. Overall, due to the sorting algorithm in both of them, DispEn with sorting method and PerEn cannot detect spikes. In contrast, DispEn and FDispEn with logsig can detect outliers, showing an advantage of the DispEn and FDispEn with logsig over sorting-based entropy methods, such as PerEn.

\section{Applications of DispEn and FDispEn to Biomedical Time Series}
In this Section, to evaluate the DispEn and FDispEn methods, compared with PerEn, to quantify the degree of the irregularity or uncertainty of biomedical signals, we use two publicly-available datasets from http://www.physionet.org. 

\subsection{Blood Pressure in Rats}
We evaluate the ability of entropy methods on the non-invasive blood pressure signals from nine salt-sensitive hypertensive (SS) Dahl rats and six rats protected (SP) from high-salt-induced hypertension (SSBN13) on a high-salt diet ($8\%$ salt) for 2 weeks \cite{goldberger2000components,fares2016effect}. Each blood pressure signal was recorded using radiotelemetry for two minutes with sampling frequency of 100 Hz. The study was approved by the Institutional Animal Care and Use Committee of the Medical College of Wisconsin-Madison, US \cite{goldberger2000components,fares2016effect}. Further information can be found in \cite{goldberger2000components,fares2016effect}.

\begin{figure*}
	\centering
	\includegraphics[width=19cm,height=4cm]{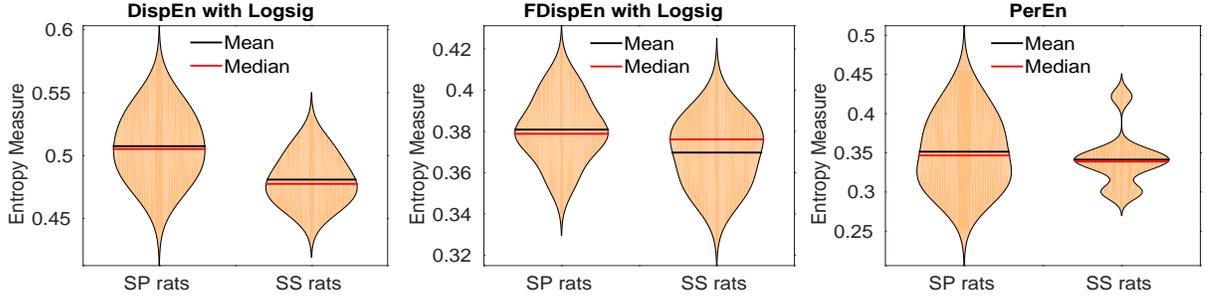}
	\caption{Mean, median, and SD of results obtained by PerEn, and DispEn and FDispEn with logsig from salt-sensitive (SS) vs. salt protected (SP) rats' blood pressure signals.}
	\label{SampEn_Blood_Pressure_Rats}
\end{figure*}

The results, illustrated in Figure~\ref{SampEn_Blood_Pressure_Rats}, show a loss of irregularity with the salt-sensitive rats, in agreement with \cite{fares2016effect}. We set $m=6$ for PerEn \cite{kowalski2007bandt}, and $m=5$ for both DispEn and FDispEn. The Hedges' g effect size was employed to assess the differences between results for SS versus SSBN13 Dahl rats. The effect sizes for DispEn, FDispEn, and PerEn are respectively 1.285 (very large difference), 0.743 (moderate difference), and 0.253 (small difference), showing an advantage of DispEn and FDispEn over PerEn, to distinguish the salt-sensitive from salt protected rats' recordings.

\subsection{Gait Maturation Database}

We also used the gait maturation database to assess the entropy methods to distinguish the effect of age on the intrinsic stride-to-stride dynamics \cite{hausdorff1999maturation}. A subset including 23 healthy boys and girls is considered in this study. The children were classified into two age groups: 3 and 4 years old (11 subjects) and 11 to 14 years old children (12 subjects). Height and weight of the young and elderly groups were $105\pm2$ cm and $155\pm10$ cm, and $17.3\pm0.7$ kg, and $44.4\pm2.7$ kg, respectively. Subjects walked at their normal pace for about 8 minutes with sampling frequency of 300 Hz. For more information, please see \cite{hausdorff1999maturation}.

\begin{figure*}
	\centering
	\includegraphics[width=19cm,height=4cm]{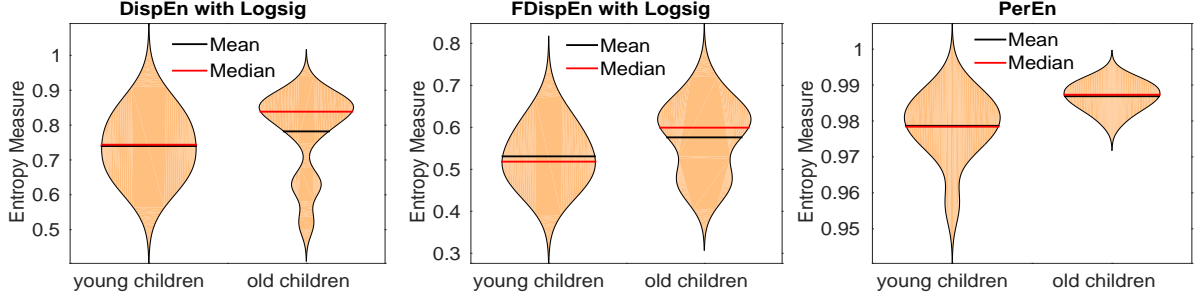}
	\caption{Mean, median, and SD of results obtained by PerEn, and DispEn and FDispEn with logsig for young and elderly children's stride-to-stride recordings.}
	\label{SampEn_Gait_Maturation_Database}
\end{figure*}

The results, depicted in Figure~\ref{SampEn_Gait_Maturation_Database}, show that the average entropy values obtained by DispEn and FDispEn with logsig, and PerEn for the elderly children are larger than those for the young children, in agreement with previous studies \cite{hong2008age,bisi2016complexity}. We set $m=4$ for PerEn \cite{kowalski2007bandt}, $m=2$ for DispEn, and $m=3$ for FDispEn. The effect size values for DispEn, FDispEn, and PerEn are respectively 0.768, 0.772, and 0.651, showing an advantage of FDispEn over FDispEn, and both DispEn and DisPEn over PerEn, to distinguish young from elderly children's signals. 

Overall, in addition to the superiority of DispEn and FDispEn over PerEn in terms of running time for long signals, the results for the real datasets evidence the advantage of DispEn and FDispEn with logsig over PerEn to distinguish different types of dynamics of biomedical recordings. In spite of the promising findings and results for different applications aforementioned in this pilot study, further investigations on potential applications of DispEn and FDispEn are recommended.

\section{Conclusions}
In this paper, we carried out an investigation aimed at gaining a better understanding of our recently developed DispEn, especially regarding the parameters and mapping techniques used in DispEn. We also introduced FDispEn to quantify the irregularity of time series in this article. The basis of the technique lies in taking into account only the frequency of signals. Although both PerEn and FDispEn are based on the frequency of signals, the latter addresses the problem of equal values existed in PerEn. The concepts of forbidden amplitude- and frequency-based dispersion patterns are introduced in this study.

The work done here has the following implications for irregularity estimation. Firstly, we showed that DispEn and FDispEn with logsig are appropriate approaches when dealing with noise. Interestingly, DispEn and FDispEn with logsig, unlike PerEn, detected outliers. We also found that the forbidden amplitude- and frequency-based dispersion patterns are suitable to distinguish deterministic from stochastic time series. DispEn and FDispEn were noticeably faster than PerEn for long time series. Finally, the results showed that both DispEn and FDispEn with logsig distinguish various physiological states of the biomedical time series better than PerEn.  

Due to their quickness and ability to detect dynamics of signals, we hope DispEn and FDispEn can be used for analysis of a wide range of physiological and even non-physiological signals.

\bibliography{RefDisEn}

\subsection*{Appendix A: SampEn vs. DispEn and FDispEn}
Sample entropy (SampEn) denotes the negative natural logarithm of the conditional probability that two series similar for \textit{m} sample points remain similar at the next sample, where self-matches are not considered in calculating the probability \cite{richman2000physiological}. For detailed information, please refer to \cite{richman2000physiological}.

In spite of its power to detect dynamics of signals, SampEn has two key deficiencies. First, SampEn values for short signals are either undefined or unreliable, as in its algorithm, the number of matches whose differences are smaller than a defined threshold is counted. When the time series length is too small, this number may be 0, leading to undefined values. Second, SampEn is not fast enough for real time applications and has a computation cost of O($N^2$) \cite{jiang2011fast}. In contrast, DispEn and FDispEn do not lead to undefined values and their computation cost is O(\textit{N}) \cite{rostaghi2016dispersion}.  

The dependence of the number of classes \textit{c} (DispEn and FDispEn) and threshold \textit{r} (SampEn) is inspected by the use of a MIX process evolving from randomness to periodic oscillations as follows \cite{escudero2009interpretation,pincus1991approximate}:
\begin{equation}
	MIX=(1-z)x+zy
\end{equation}    
where \textit{z} is a random variable which equals to 1 with probability \textit{p} and equals to 0 with probability $1-p$, \textit{x} denotes a periodic synthetic time series created by $x_k=\sqrt{2}\sin(\frac{2\pi k}{12})$, and \textit{y} is a uniformly distributed variable on $[-\sqrt{3},\sqrt{3}]$ \cite{escudero2009interpretation,pincus1991approximate}. The time series was based on a MIX process whose parameter linearly varied between 0.99 and 0.01. Therefore, this series evolved from randomness to orderliness. The signal has a sampling frequency of 150  Hz and a length of 100 s (15000 samples). The techniques are applied to the 20 realizations of the MIX process using a moving window of 1500 samples (10 s) with $50\%$ overlap. We used different threshold values $r=0.1, 0.2, 0.3, 0.4$, and 0.5 of SD of the signal \cite{richman2000physiological} for SampEn, and $c=2, 4, 6, 8$ and 10 for DispEn and FDispEn. 

\begin{figure*}
	\centering
	\includegraphics[width=16cm,height=5cm]{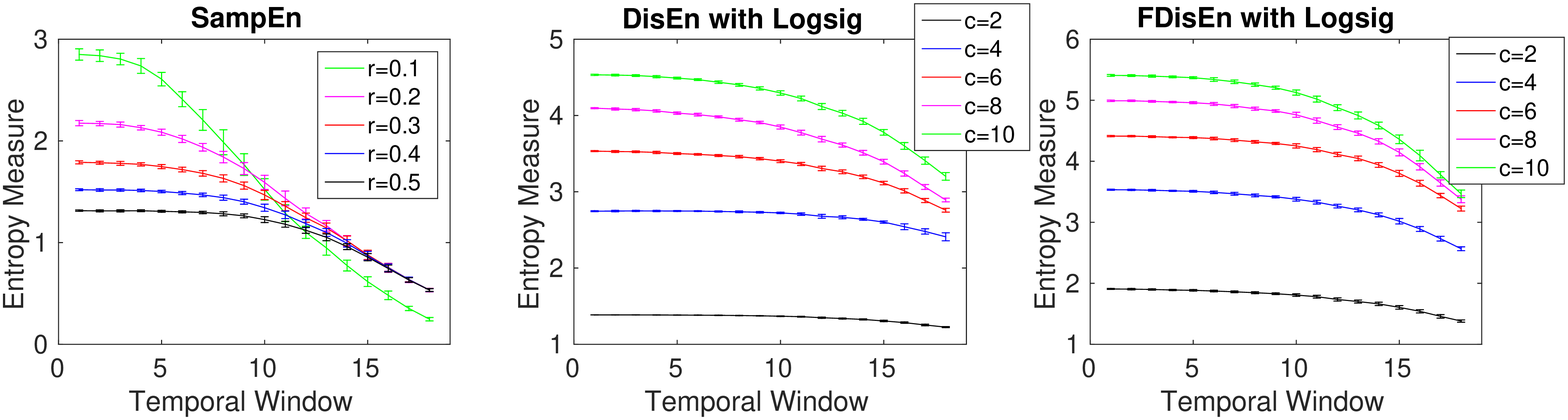}
	\caption{(a) Average and (b) SD of entropy values obtained by the DispEn, FDispEn, and SampEn with different number of classes (for DispEn and FDispEn) and different threshold values (SampEn) using a MIX process evolving from randomness to periodic oscillations. We used a window with length 1500 samples moving along the MIX process (temporal window).}
	\label{Sensivity_Threshold_Class_DisEn_SampEn}
\end{figure*}

The results, depicted in Figure~\ref{Sensivity_Threshold_Class_DisEn_SampEn}, show that the mean entropy values using all these approaches are the least in higher temporal windows, in agreement with the previous studies \cite{escudero2009interpretation,pincus1991approximate}. The results also evidence that the number of classes (\textit{c}) in DispEn and FDispEn is inversely related to the threshold value \textit{r} used in the SampEn algorithm. It is worth noting that SampEn, unlike DispEn and FDispEn, is not consistent as $r=0.1$ crosses the other lines. We set $m=2$, 2, and 3, for respectively SampEn, DispEn, and FDispEn, as recommended before. 

To compare the results obtained by the entropy algorithms, we used the coefficient of variation (CV) defined as the SD divided by the mean. We use such a metric as the SDs of signals may increase or decrease proportionally to the mean. We inspect the MIX process with length 1500 samples and $p=0.5$ as a trade-off between random ($p=1$) and periodic oscillations ($p=0$). The CV values, depicted in Table~\ref{Results_CV_Noise_MIX}, show that DispEn- and FDispEn-based CV values for different number of classes are noticeably smaller than those for SampEn with different threshold values, showing an advantage of DispEn and FDispEn over SampEn.

\begin{table*}
	\centering
	\setlength{\tabcolsep}{5pt}
	\label{tab:table4}\caption{CVs of DispEn and FDispEn with logsig, and PerEn values for the MIX process with $p=0.5$ and length 1000 samples.} 
	\begin{tabular}{c*{8}{c}}
		\hline
		DispEn ($c\!=\!2$) & DispEn ($c\!=\!4$)& DispEn ($c\!=\!6$)&DispEn ($c\!=\!8$)&DispEn ($c\!=\!10$)\\
		\hline
		0.0021& 0.0034 & 0.0045 & 0.0041  & 0.0048  \\
		\\
		FDispEn ($c\!=\!2$)  & FDispEn ($c\!=\!4$)& FDispEn ($c\!=\!6$)&FDispEn ($c\!=\!8$)&FDispEn ($c\!=\!10$)\\
		\hline
		0.0078& 0.0064 & 0.0040 &  0.0043 & 0.0049 \\
		\\
		\hline
		SampEn ($r\!=\!0.1\times$SD)  & SampEn ($r\!=\!0.2\times$SD)& SampEn ($r\!=\!0.3\times$SD)&SampEn ($r\!=\!0.4\times$SD)&SampEn ($r\!=\!0.5\times$SD)\\
		\hline
		0.0604 & 0.0342&
		0.0224 & 0.0174
		& 0.0150     \\

	\end{tabular}
	\label{Results_CV_Noise_MIX}
\end{table*}

\end{document}